\documentclass[acmsmall]{acmart}

\AtBeginDocument{%
  \providecommand\BibTeX{{%
    \normalfont B\kern-0.5em{\scshape i\kern-0.25em b}\kern-0.8em\TeX}}}

\setcopyright{acmlicensed}
\acmJournal{PACMMOD}
\acmYear{2023} \acmVolume{1} \acmNumber{1} \acmArticle{110} \acmMonth{5} \acmPrice{15.00}\acmDOI{10.1145/3588964}

\usepackage{microtype}
\usepackage{graphicx}
\usepackage{subfigure}
\usepackage{booktabs} 

\usepackage{hyperref}
\usepackage{xcolor}
\usepackage[linesnumbered,ruled,noend]{algorithm2e}
\usepackage{multirow}
\usepackage{listings}
\usepackage{epstopdf}
\usepackage{iitem}
\usepackage{framed}
\usepackage{setspace}
\usepackage{verbatim}
\usepackage{enumitem}
\usepackage[misc]{ifsym}

\newcommand{\oursys}{{\textsc{FlexMoE}}\xspace}
\newcommand{\vexpert}{{\textsc{vExpert}}\xspace}

\DeclareMathOperator*{\argmax}{arg\,max}
\DeclareMathOperator*{\argmin}{arg\,min}

\begin{document}

\title{\oursys{}: Scaling Large-scale Sparse Pre-trained Model Training via Dynamic Device Placement}

\author{Xiaonan Nie}
\email{xiaonan.nie@pku.edu.cn}
\affiliation{%
\department{School of CS \& Key Lab of High Confidence Software  Technologies (MOE)}
  \institution{Peking University}
  \city{Beijing}
  \country{China}
}

\author{Xupeng Miao}
\email{xupeng@cmu.edu}
\affiliation{%
  \institution{Carnegie Mellon University}
  \country{USA}
}

\author{Zilong Wang}
\email{zilongwang@microsoft.com}
\affiliation{%
  \institution{Microsoft}
  \country{China}
}

\author{Zichao Yang}
\email{yangtze2301@gmail.com}
\affiliation{%
  \institution{Carnegie Mellon University}
  \country{USA}
}

\author{Jilong Xue}
\email{jxue@microsoft.com}
\affiliation{%
  \institution{Microsoft Research}
  \country{China}
}

\author{Lingxiao Ma}
\email{lingxiao.ma@microsoft.com}
\affiliation{%
  \institution{Microsoft Research}
  \country{China}
}

\author{Gang Cao}
\email{caogang@baai.ac.cn}
\affiliation{%
  \institution{Beijing Academy of Artificial Intelligence}
  \country{China}
}

\author{Bin Cui}
\authornote{Bin Cui is the corresponding author.}
\email{bin.cui@pku.edu.cn}
\affiliation{%
  \department{School of CS \& Key Lab of High Confidence Software  Technologies (MOE), Institute of Computational Social Science, Peking University (Qingdao)}
  \institution{Peking University}
  \city{Beijing}
  \country{China}
}

\renewcommand{\shortauthors}{Xiaonan Nie, et al.}

\begin{abstract}
With the increasing data volume, there is a trend of using large-scale pre-trained models to store the knowledge into an enormous number of model parameters. The training of these models is composed of lots of dense algebras, requiring a huge amount of hardware resources. Recently, sparsely-gated Mixture-of-Experts (MoEs) are becoming more popular and have demonstrated impressive pretraining scalability in various downstream tasks. However, such a sparse conditional computation may not be effective as expected in practical systems due to the routing imbalance and fluctuation problems. Generally, MoEs are becoming a new data analytics paradigm in the data life cycle and suffering from unique challenges at scales, complexities, and granularities never before possible.

In this paper, we propose a novel DNN training framework, \oursys{}, which systematically and transparently address the inefficiency caused by dynamic dataflow.
We first present an empirical analysis on the problems and opportunities of training MoE models, which motivates us to overcome the routing imbalance and fluctuation problems by a dynamic expert management and device placement mechanism.
Then we introduce a novel scheduling module over the existing DNN runtime to monitor the data flow, make the scheduling plans, and dynamically adjust the model-to-hardware mapping guided by the real-time data traffic.
A simple but efficient heuristic algorithm is exploited to dynamically optimize the device placement during training. 
We have conducted experiments on both NLP models (e.g., BERT and GPT) and vision models (e.g., Swin). And results show \oursys{} can achieve superior performance compared with existing systems on real-world workloads --- \oursys{} outperforms DeepSpeed by $1.70\times$ on average and up to $2.10\times$, and outperforms FasterMoE by $1.30\times$ on average and up to $1.45\times$. 
\end{abstract}


\received{July 2022}
\received[revised]{October 2022}
\received[accepted]{November 2022}

\maketitle

\section{Introduction}
\label{sec:intro}
Large-scale pre-trained models (PTMs), such as Transformer models, have promoted the deep learning (DL) development on various complicated tasks, including natural language processing, e.g., BERT~\cite{bert}, GPT~\cite{gpt3}, T5~\cite{t5}, computer vision, e.g., ViT~\cite{vit}, Swin~\cite{liu2021swin}, advertising recommendation, e.g., M6~\cite{lin2021m6}, and so on. These models are also known as foundation models since they are trained on hundreds of gigabytes of data and can be adapted, e.g., task-specific fine-tuning, to a wide range of downstream tasks. 
However, recent studies~\cite{scalinglaws,gpt3} have demonstrated the model quality scales as the power law with data size, parameter size, and computation budgets. Current state-of-the-art foundation models have excessively grown to trillions of parameters and therefore become extremely expensive and time-consuming to be trained.

To avoid hitting the train-ability wall, one well-known line of studies utilizes the sparsely-gated Mixture-of-Experts (MoE)~\cite{shazeer2017outrageously} structure to make pre-trained models cost-effective.
This sparse MoE architecture enlarges the model parameter size by expanding expert networks while keeping the computational budget stable. Specifically, each MoE layer consists of a gate network and a series of experts (e.g., tens of), where the gate network routes the input tokens (i.e., the inputs of the layer) to only a small number of experts rather than all experts to be computation-efficient. 
MoE models have been successfully used to boost the performance of various PTMs, such as language models (e.g., GLaM~\cite{du2021glam} achieve the state-of-the-art performance in the SuperGLUE NLP benchmark~\cite{DBLP:conf/nips/WangPNSMHLB19}) and vision models (e.g., V-MoE~\cite{vmoe} matches the most powerful vision transformers with only half computation time).

Due to the superior behaviors, large-scale MoE models have attracted wide interests in industrial companies recently when building real-world big data applications, including Google~\cite{du2021glam}, Meta~\cite{HashLayer}, Microsoft~\cite{rajbhandari2022deepspeed} and Alibaba~\cite{lin2021m6}.
Such kind of data-intensive workloads is becoming a new data analytics paradigm for data science research communities. With the rapid expansion of MoE models, unprecedented opportunities are brought into the data life cycle (especially in model development, deployment, and management procedures), as well as challenges. 
For example, Google created the ST-MoE~\cite{stmoe} model to help improve the question answering quality of their search engines, which achieves state-of-the-art result in the AIRC leadboard~\cite{qa_leadboard}, e.g., increasing from $81.40\%$ to $86.52\%$ by using MoE. However, such a large MoE model not only needs a lot of data to train on, e.g., 1.5T tokens for ST-MoE, but also consumes large amounts of expensive computing resources for training. 
Therefore, it is necessary to address data management problems hidden in existing systems and conduct optimizations to overcome these challenges.

Since a single MoE layer can easily exceed the limited GPU memory, previous distributed training techniques (e.g, data parallel~\cite{torchdistributed} and model parallel~\cite{megatron-lm,narayanan2019pipedream, miao2023galvatron, nie2022tsplit}) are becoming unsatisfactory. To train large MoE models, \citet{GShard} proposed the \textit{expert parallelism} technique, which distributes experts into multiple GPUs to fit within the limited GPU memory.
For each training token, after the gate network has determined the routing, it will be sent to the GPUs where its target experts reside.
Since it adapts to the special characteristic of MoE structures, expert parallelism has become one of the most widely-used techniques to train MoE-based models in several well-known frameworks, including DeepSpeed-MoE~\cite{rajbhandari2022deepspeed}, Tutel~\cite{tutel}
and HetuMoE~\cite{nie2022hetumoe}.

\textbf{The Challenge of Workload Imbalance.} Despite the widespread adoption of expert parallelism~\cite{GShard}, it suffers from the severe workload imbalance brought by the sparse and conditional computing nature of MoE layers. 
Specifically, the input tokens are organized at a fine-grained level and routed to the individual experts by the gating network. 
We found that the input tokens show highly uneven preferences on the experts and the routing results keep varying over training~\cite{densetosparse}. As reported by ~\cite{BaseLayer,he2022fastermoe,ma2022bagualu} and verified by our empirical studies, such workload imbalance could lead to significantly efficiency slowdown and severe resource under-utilization.

\textbf{The Current Landscape.} To tackle the workload imbalance problem, most existing works leverage an auxiliary balance loss on each MoE layer to enforce balanced routing. To control the trade-off between system efficiency (workloads balance) and statistical efficiency (model quality), the balance loss is further associated with a tunable penalty weight. In addition, ~\citet{fedus2021switch} introduces the expert capacity to avoid excessive input token assignments to the experts. In other words, tokens beyond the expert capacity will be dropped by the experts. Nevertheless, these methods can only mitigate the workload imbalance problem rather than fundamentally guarantee balance. Moreover, they make users stuck in a dilemma: \textit{whether and to what degree shall we sacrifice model quality for system efficiency}? For instance, applying a large penalty weight to the balance loss (or applying a low capacity to experts) could help load balancing among experts but harm the model quality since a substantial amount of tokens would be routed to some less-desired experts (or dropped), and vice versa.

Intuitively speaking, the aforementioned methods try to adjust the data-flow of tokens to be as even as possible, i.e., enforcing all experts to process almost the same amount of tokens. They are friendly adapted to existing expert parallelism systems, because users can easily change the definitions of their model (e.g., balance loss and capacity) and without any modifications on the implementation of underlying DL systems. In other words, such approaches are friendly to people who are not familiar with DL systems so they have to sacrifice the model quality for training efficiency.
Unlike previous approaches making model modifications, which may cause model quality degradation, we try to optimize the workload imbalance from a system perspective:

\begin{quote}
\em
How to design a distributed MoE training system that could achieve the maximum system efficiency without affecting the model quality in the meanwhile?
\end{quote}

\textbf{Summary of Our Approach.} In this work, we propose \oursys{}, a novel distributed training system for large-scale sparse MoE models.
We overcome the \textbf{routing imbalance} problem from the brand new perspective of the expert placement --- 
instead of placing each expert on a single GPU device as in traditional expert parallelism nor duplicating all experts to all GPU devices as in traditional data parallelism, we introduce a \textbf{fine-grained replicated expert parallelism} that selects specific heavy experts, duplicates them over multiple devices, and spreads the input tokens across the replicas.
It is non-trivial in sparse MoE models because the device placement of these experts changes the original input data assignment (i.e., all-to-all communication across different experts) and involves additional synchronization overheads (i.e., all-reduce communication among replicated experts). 
By precisely modeling the actual execution procedure of the MoE layer, \oursys{} estimates the costs that determine the concrete expert replication scheme. 

We further investigate the uneven distribution of the sparsely-activated experts in various MoE training workloads. 
Our key finding is that MoE models exhibit \textbf{routing fluctuation} during the iterative training process. 
We observe that the imbalanced expert preference distribution changes continuously and smoothly during the training process (will be described in-depth in Section~\ref{sec:observation}).
The training of gate networks is highly unstable~\cite{dai2022stablemoe} because gradient-based optimization algorithms will reinforce the previous routing determination until it reaches a certain point and escapes from the wrong learning direction. To handle this obstacle, we design a \textbf{dynamic expert management} mechanism in \oursys{} to adaptively adjust the expert-to-device mapping and the assignment of tokens during training.
Specifically, we adopt a \textit{data-driven approach} to adaptively change the expert placement during training by monitoring the traffic trends. 
For instance, \oursys{} gradually expands resources for experts with increasing workloads for faster calculations and shrinks those with decreasing workloads to reduce replica synchronization overheads.

\oursys{} is designed as 
a novel scheduling module
on top of existing DNN frameworks, which monitors data traffic, makes scheduling plans, and dynamically adjusts the mapping between the data-flow graph and distributed devices. 
Particularly, three placement adjustment primitives (i.e., expand, shrink, migrate) are provided to flexibly govern the expert placement and a gate flow-control mechanism is introduced to enable autonomous global traffic optimization. 
Based on these building blocks, a simple yet effective algorithm is designed to estimate the benefits or overheads of each adjustment to determine the optimal scheduling plan. 

Last but not least, we implement \oursys{} on top of PyTorch~\cite{pytorch19} and verify the superiority of our work through comprehensive experiments. When training several popular MoE-based PTMs (e.g., BERT, GPT, Swin) of various tasks on a 64-GPU A100 cluster, \oursys{} achieves up to 2.1$\times$ of speedup compared with existing state-of-the-art MoE training systems. Furthermore, \oursys{} consistently outperforms them in terms of model quality since we do not accommodate the expert placement with compromised token-routing.

\textbf{Paper Organization.} 
In section~\ref{sec:background}, we discuss the inefficiency of training sparse MoE models with expert parallelism and explore opportunities for optimizing such problems through system-level optimizations.
In section~\ref{sec:design}, we introduce the system design of \oursys{}, which leverages the vExpert abstraction to facilitate dynamic expert management and device placement.
We provide details on the system implementation in Section~\ref{sec:impl} and evaluate the effectiveness of \oursys{} in Section~\ref{sec:eval}.

\label{sec:intro}

\section{Background and Motivations}
\label{sec:background}
We first present the notations used in our paper.

\begin{itemize}[leftmargin=15pt]
    \item[*] $\mathcal{E}$: A series of experts $\{e_1,...,e_N \}$.
    \item[*] $\mathcal{G}$: A set of GPUs, where $g \in \mathcal{G}$.
    {\item[*] $\mathcal{I}$: The assignment of tokens, where $I_{e,g} \subset \mathcal{I}$.\\ $I_{e,g}$ represents the number of tokens for expert $e$ to GPU $g$.}
    {\item[*] $\mathcal{P}$: The expert-to-device mapping, where $(e,g) \in \mathcal{P}$. \\
    $(e,g) \in \mathcal{P}$ represents GPU $g$ exists expert $e$.}
    \item[*] $TPS$: Tokens-per-second for an expert.
    \item[*] $Bw_{g, g'}$: Bandwidth between GPU $g$ and $g'$.
    {\item[*] $BPS(\mathcal{G}')$: Bytes-per-second for AllReduce on a set of GPUs $\mathcal{G}'$.}
\end{itemize}

\subsection{Transformer with Mixture-of-Expert} 
\label{sec:transformer_and_moe}
The Transformer architecture~\cite{DBLP:conf/nips/transformer} has demonstrated its superior performance in various tasks~\cite{bert, gpt3, t5, vit, liu2021swin, jcst_22, dase22}, which mainly consists of attention networks and feed-forward networks.  
Each attention network first linearly transforms the input tokens into corresponding queries (Q), keys (K) and values (V) and then performs the scaled dot-product on them as Equation~\ref{equ:attention}, where $d$ is the dimension of queries and keys.
Meanwhile, this attention network benefits from capturing the dependencies between tokens within the sequence, and thus is effective in sequence transduction tasks. 
\begin{equation}
    \texttt{Attention}(Q, K, V) = \texttt{softmax}(\frac{QK^T}{\sqrt{d}})V
\label{equ:attention}
\end{equation}
Each feed-forward network (FFN) is composed by two fully connected layers and an activation function, i.e., ReLU, formulated as Equation~\ref{equ:ffn}. Different from the attention network, FFNs model the relation of different feature dimensions within a single token by scaling it into larger dimension space. For example, the output dimension of $W_1$ is always set as $4\times$ of its input dimension. Existing famous pre-trained models are usually stacked by a series of transformer layers to improve its model capacity as well as model quality.
\begin{equation}
    \texttt{FFN}(x) = W_{2} \cdot \texttt{ReLU}(W_{1} \cdot x + b_{1}) + b_{2}
\label{equ:ffn}
\end{equation}


Recently, scaling with more data and more parameters has driven significant model quality improvement~\cite{vit, gpt3} while requires large amounts of computing resources for training.
To scale models efficiently, researchers have adopted the sparse-gated Mixture-of-Experts (MoE) paradigm~\cite{shazeer2017outrageously, fedus2021switch} by replacing the FFN network with an MoE layer, where each input token activates only a subset of model parameters, and thus introducing model sparsity. 

\begin{figure}[t]
    \centering
    \subfigure{
        \includegraphics[width=0.1\columnwidth]{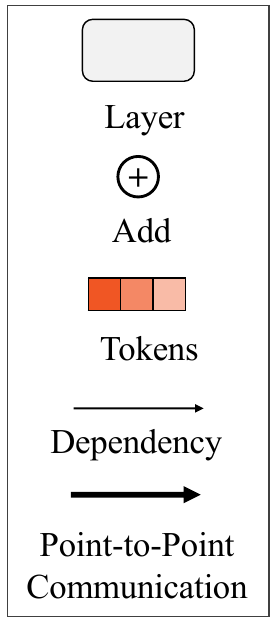}
    }
    \setcounter{subfigure}{0}
    \subfigure[The workflow of an MoE layer]{
        \includegraphics[width=0.28\columnwidth]{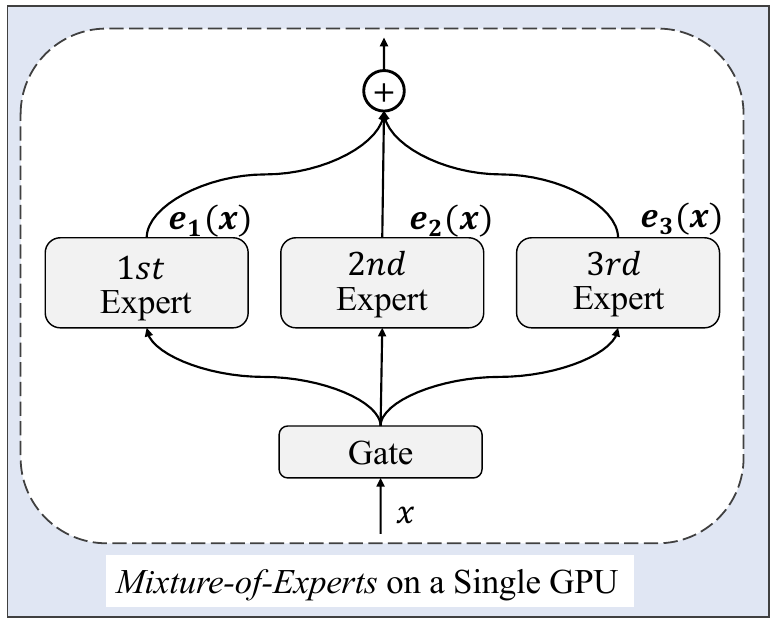}
        \label{fig:moe_layer}
    }
    \subfigure[The distributed training of an MoE layer with Top-1 gate]{
        \includegraphics[width=0.49\columnwidth]{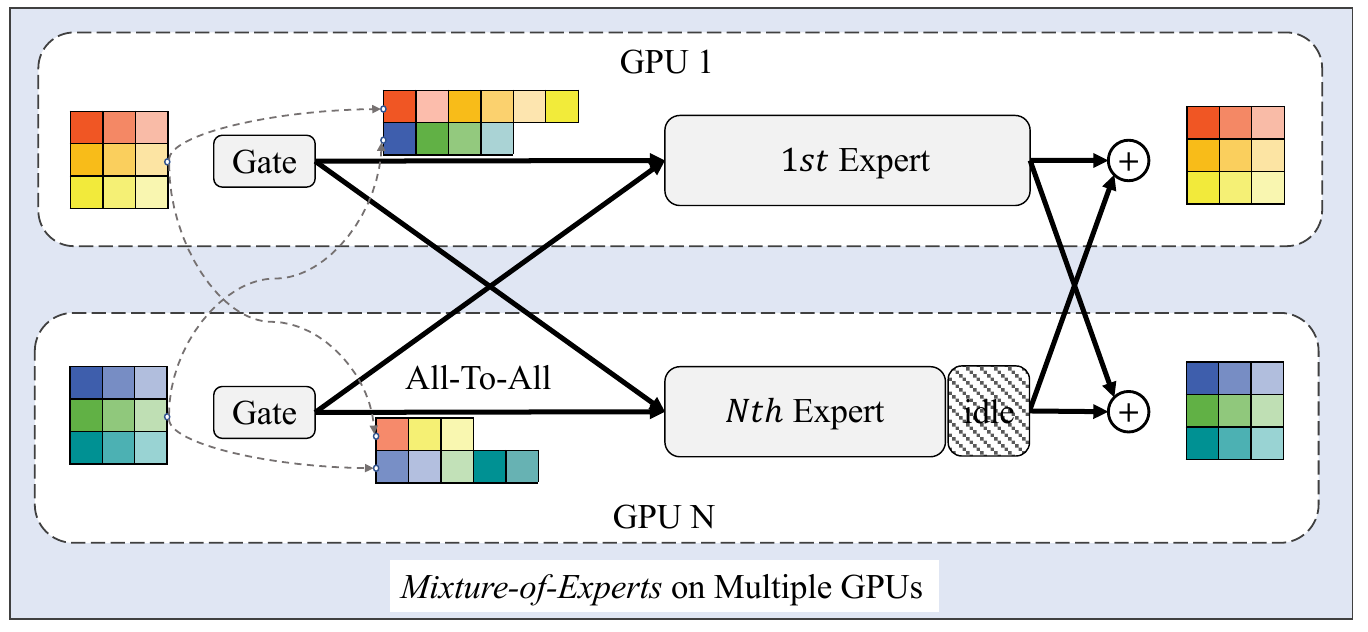}
        \label{fig:moe_layer_distributed}
    }
    \vspace{-3mm}
    \caption{Illustrations of Mixture-of-Experts (MoE). Figure~\ref{fig:moe_layer} represents the workflow of a typical MoE layer (detailed in Section~\ref{sec:transformer_and_moe}). Figure~\ref{fig:moe_layer_distributed} presents the distributed training of an MoE layer as expert parallelism, where experts are partitioned while non-MoE layers (e.g., self-attention, gate) are duplicated across devices. For example, according to the result of \texttt{Gate}, GPU 1 routes 6 input tokens to $1st$ Expert and the other 3 tokens to $Nth$ Expert (detailed in Section~\ref{sec:distributed_training}).}
    \vspace{-4mm}
    \label{fig:background}
\end{figure}

The key components of a MoE layer include a data-dependent sparse gate network $g(x)$ and a series of experts $\mathcal{E}$, as shown in Figure~\ref{fig:moe_layer}. 
For each input token $x$, the gate network first produces the probability of $x$ with respective to all experts and then routes $x$ to its corresponding experts. 
The Top-K gate~\cite{shazeer2017outrageously} is formulated in Equation~\ref{equ:topk_gate}, which keeps only the top k values before the softmax function.
\begin{equation}
    g(x) = \texttt{softmax}(\texttt{TopK}(x \cdot W_{g}) )
\label{equ:topk_gate}
\end{equation}

As soon as each expert $e_i$ receives its input token $x$ ($e_i \in \mathcal{E}$), it produces its corresponding output $e_i(x)$ and then the final output $y$ is obtained by the linear combination of $e_i(x)$ weighted by the output of the gate network  $g(x)_i$ as follows:
\begin{equation}
    y = \sum_{i=1}^{N}g(x)_{i} \cdot e_{i}(x).
\label{equ:moe_weighted_sum}
\end{equation}

When the gate network's sparsity is static, e.g., Top-1~\cite{fedus2021switch} or Top-2~\cite{GShard}, the computation and communication costs of given inputs nearly remain constant as the number of experts increases, allowing models to be scaled effectively with MoE.

To help readers better understand the routine of MoE layers, we also provide an illustration in Figure~\ref{fig:moe_layer_distributed}, where each GPU is associated with one expert and 9 tokens are fed to the gate network and routed to their target experts. For example, GPU 1 routes 6 and 3 tokens to the $1st$ and $Nth$ experts, while GPU N routes 4 and 5 tokens to these two experts. This assignment would cause the workload imbalance problem under current expert placement (i.e., 10 tokens for GPU 1 and 8 tokens for GPU N). 
Since there is a reverse routing step after the expert computation, all GPUs have to wait for each other before executing the following layers. Such an imbalanced workload inevitably results in GPU under-utilization and low training efficiency.

\subsection{Distributed Training}
\label{sec:distributed_training}
As models are sparsely scaled with MoE, multiple GPUs would be involved for model management and training acceleration~\cite{megatron-lm, rajbhandari2022deepspeed, nie2023angel}.
We will analyze the existing popular parallelism strategies in the following:

\textbf{Data Parallelism.} 
Data parallelism (DP) is usually used to scale training when the model can fit in the device's available GPU memory, as the communication primitives (e.g., AllReduce) of DP achieve good scalability performance.
In DP, training samples are partitioned while model parameters are duplicated for each device. Given a batch of training samples, each worker executes the forward and backward computation, synchronizes gradients globally (i.e., averaged), and updates local parameters based on the synchronized gradients. However, each device has to maintain a full replica of the model and thus DP can't be used to scale up to large models.

\textbf{Model Parallelism.} If the memory requirement of the given model exceeds the GPU memory, it should be partitioned across multiple devices.
In tensor parallelism (TP), every single tensor can be partitioned over multiple devices to reduce the memory footprint of each GPU. For example, Megatron-LM~\cite{megatron-lm} proposed to partition the attention network by exploiting its inherent parallelism in the multi-head attention operation where the queries ($Q$), keys ($K$) and values ($V$) matrices can be partitioned in a column-parallel fashion. 
Pipeline parallelism (PP) is an alternative partitioning method, where different groups of layers are placed on different devices. For example, GPipe~\cite{DBLP:conf/nips/Gpipe} partitions the input batch into a number of micro batches and pipelines each device's computation across micro batches to improve the resource utilization of devices.

\textbf{Expert Parallelism.} GShard~\cite{GShard} first proposed expert parallelism (EP) as a specific model parallelism method for MoE models, where experts within an MoE layer are placed on different devices and non-MoE layers are replicated on devices as DP. 
The workflow of distributed training a Top-1 gate MoE layer is illustrated in Figure~\ref{fig:moe_layer_distributed}. After getting the target expert for each token, an All-to-All communication is involved to send tokens to target experts for processing. And another All-to-All communication is needed to send data back for the execution of data-parallel non-MoE layers.
As MoE models often have numerous experts, e.g., 1024 Experts in Switch Transformer~\cite{fedus2021switch}, EP can scale up with model size better than model parallelism.




\subsection{Problem Formulation} 
To model the training cost, we consider a single MoE layer, which consists of a gate network and a set of experts $\mathcal{E}$ ($e \in \mathcal{E}$).
Given current expert-to-device mapping $\mathcal{P}$ ($(e,g) \in \mathcal{P}$ represents GPU $g$ has expert $e$), the assignment of tokens $\mathcal{I}$ ($I_{e,g}$ represents the number of tokens for expert $e$ to GPU $g$) at the current step, the training cost can be formulated as $T(\mathcal{I}, \mathcal{P})$. Our objective is to minimize this training cost, which can be expressed as follows:
\begin{align}
\mathop{\min}\ T(\mathcal{I}, \mathcal{P}) &= \mathop{\min}\ ( {\mathop{\max}_{g \in \mathcal{G}}{\sum_{e}^{(e,g) \in \mathcal{P}}\{{{T_C(I_{e,g}) + T_{A2A}(I_{e,g})}}} + T_{Sync}(\mathcal{P}, e)}\})
\label{equ:problem_formulation}
\end{align}

\noindent The first two terms represent the cost of expert computation and All-to-All communication respectively, which are determined by the assignment of tokens, i.e., $I_{e,g}$.
And the third term represents the cost of synchronization ($T_{Sync}$) because one expert may exist on multiple GPUs as data parallelism and needs to synchronize their gradients to maintain consistent. 
Each GPU sums up all of its local experts with regard to these three terms to obtain its execution time and the global training time $T(\mathcal{I}, \mathcal{P})$ is the maximum execution time among these GPUs. 
In order to minimize the training cost while not modify the definitions of models, we design a dynamic expert management mechanism
to adaptively adjust the expert-to-device mapping $\mathcal{P}$ and the assignment of tokens $\mathcal{I}$ 
during training.

\begin{figure}[t]
  \centering
  \includegraphics[width=0.5\linewidth]{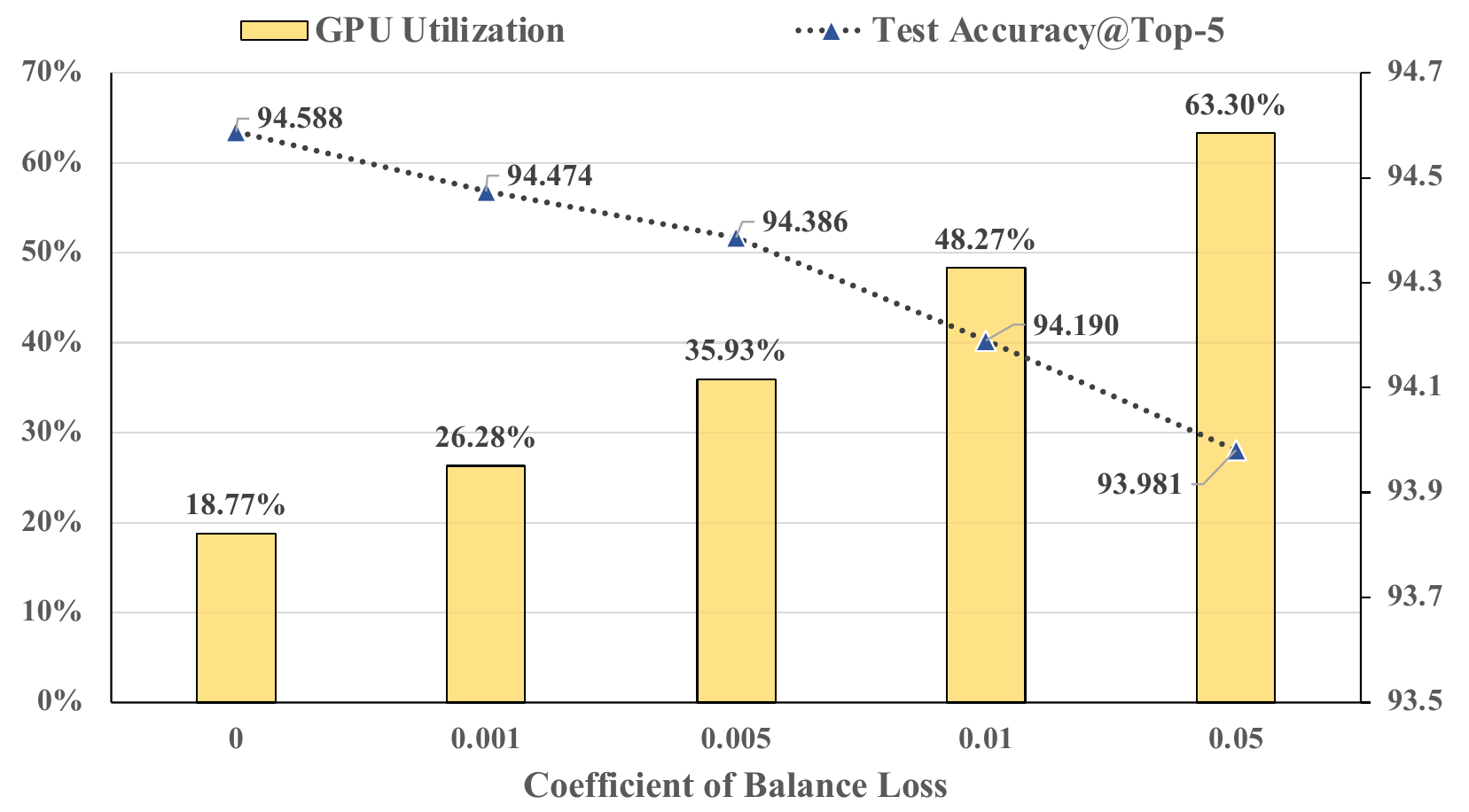}
  \caption{The top-5 accuracy of Swin-MoE under different balance loss coefficients, where we do not restrict the capacity of each expert to ensure no token is absent.}
  \vspace{-3mm}
  \label{fig:motivation}
\end{figure}

\textbf{Existing System-Friendly Optimizations.}
 As discussed in Section~\ref{sec:intro}, existing works focus on addressing the workload imbalance problem by modifying the definitions of models (e.g., balance loss and capacity), which doesn't need users to make  modifications on the underlying DL systems and therefore is system-friendly.

\begin{itemize}[leftmargin=*]
    \item {The \textit{balance loss} depicts the imbalance level of $\mathcal{I}$ and is widely used in MoE training~\cite{fedus2021switch,GShard}. Once the gate network produces an imbalanced token assignment, it would be penalized by a large balance loss. In practice, the training process aims to minimize the weighted sum of balance loss and training loss. Thus, there exists a trade-off --- emphasizing the balance loss would drive the assignment $\mathcal{I}$ to be even but harms the model quality since the training loss would be higher, and vice versa.}
    
    \item {The \textit{capacity} is a threshold that limits the number of input tokens for each expert (i.e., ${I}_{e, g}$). Tokens that exceeds the capacity will be skipped by the expert and directly forwarded to the next layer by the residual connection. In short, the capacity upper bounds the training cost of the heaviest expert to improve the overall efficiency, however, at the price of degraded model quality since a certain amount of tokens cannot be fully trained.}
\end{itemize}

\subsection{Observations and Opportunities} 
\label{sec:observation}

\textbf{Observation 1: Limitations of the existing techniques.}
We first studied the effect of the existing techniques empirically and took Swin-MoE~\cite{liu2021swin} as an example, where we varied the balance loss coefficient and did not restrict the capacity of each expert to ensure no token was absent.
The results are summarized in Figure~\ref{fig:motivation}. 
By increasing the balance loss coefficient from 0 to 0.05, the GPU utilization is improved from $18.77\%$ to $63.30\%$ while the top-5 accuracy is decreased from $94.588\%$ to $93.981\%$. 
It demonstrates that enforcing workload balance by
adjusting the assignment of tokens $\mathcal{I}$ 
would inevitably lead to the trade-off between system efficiency and model quality.

\begin{figure*}[t]
    \centering
    \setlength{\abovecaptionskip}{0cm} 
    \setlength{\belowcaptionskip}{0cm} 
    \captionsetup[subfigure]{justification=centering}
        \subfigure[CDF of expert loads]{
            \scalebox{0.6}{\includegraphics[width=0.41\columnwidth]{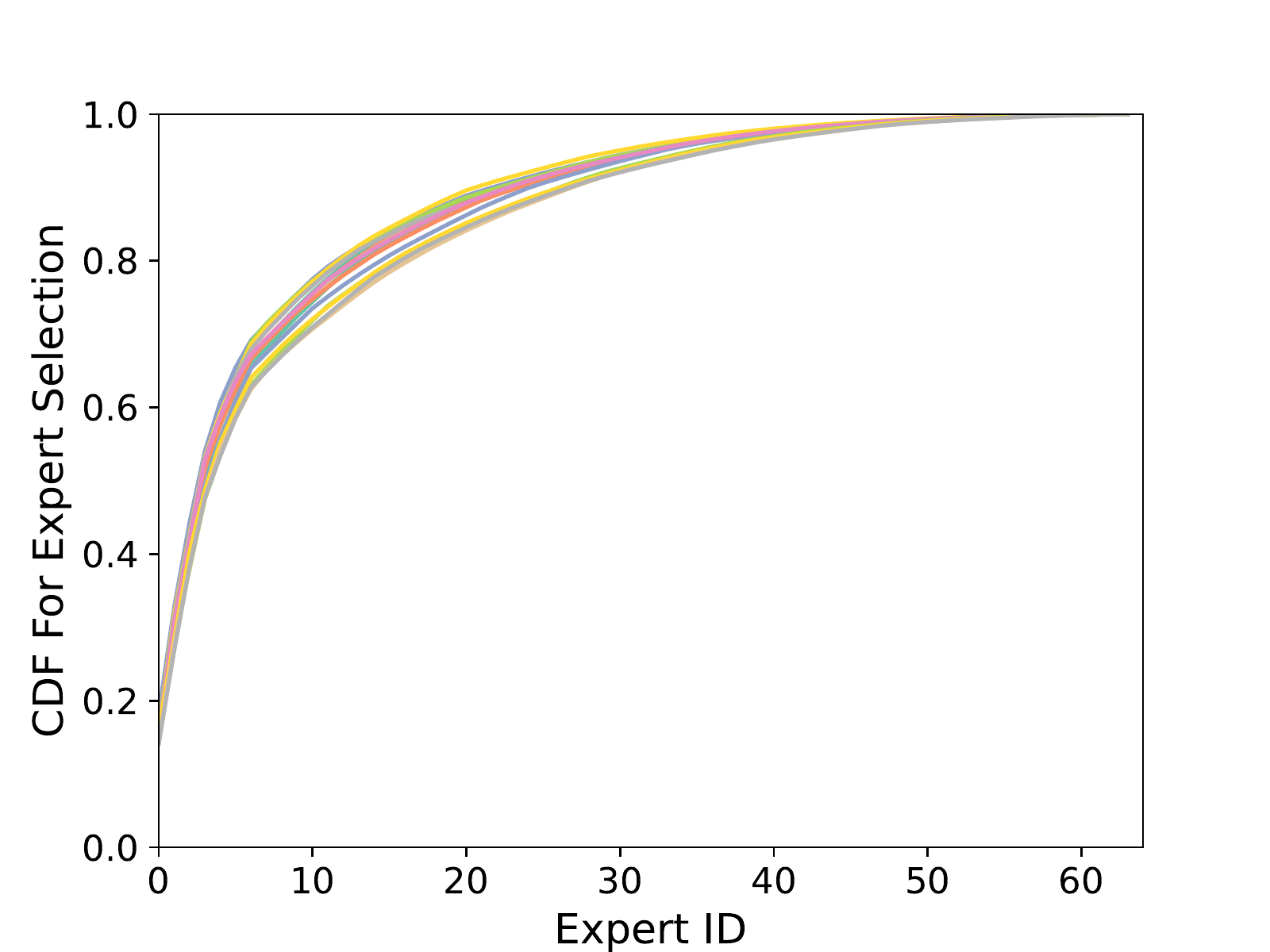}}
            \label{fig:skewness}
        }
        \subfigure[The evolution of expert loads throughout the entire training process]{
            \scalebox{0.85}{\includegraphics[width=0.75\columnwidth]{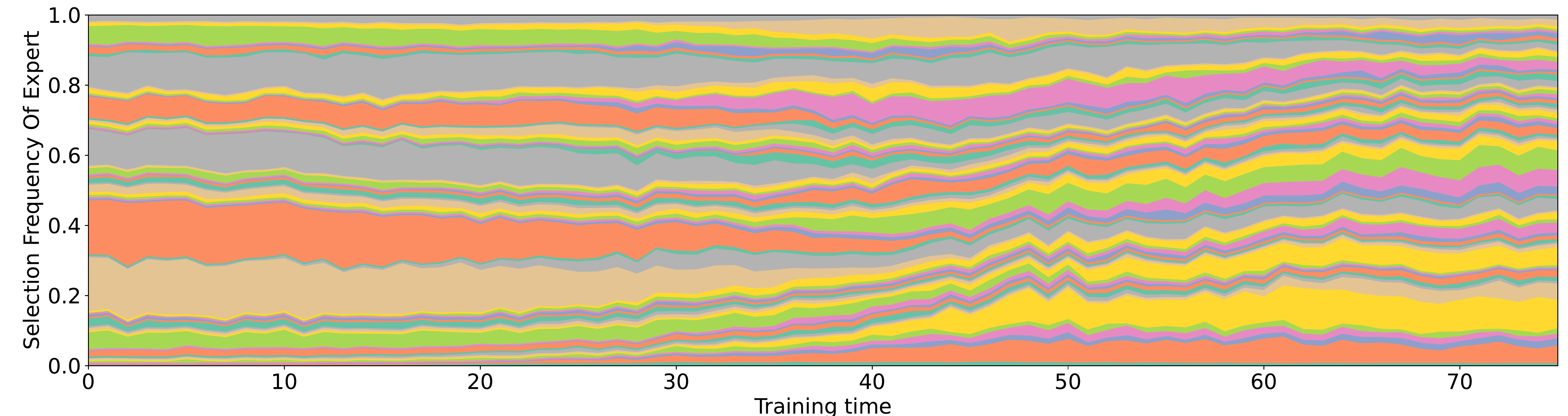}}
            \label{fig:dynamic}
        }
\caption{Illustrations of expert loads. For a single step, we sort experts within each MoE layer based on their computational loads and visualize the corresponding cumulative distribution function (CDF) in Figure~\ref{fig:skewness}, where different colors indicate different MoE layers.
For the entire training process, Figure~\ref{fig:dynamic} illustrates the dynamic changes in the load of each expert, where the different color areas represent the different experts.}
\label{fig:observation}
\vspace{-5mm}
\end{figure*}

\textbf{Observation 2: Dynamic imbalanced workloads.}
We also recorded the trace of training GPT-MoE models (64 expert per MoE layer) and studied the loads of experts during training.
Results are summarized in Figure~\ref{fig:observation} and we have identified two key characteristics:
 
\begin{itemize}[leftmargin=*]
    \item \textbf{Skewness}. As shown in Figure~\ref{fig:skewness}, we visualize the computational loads of experts as the cumulative distribution function (CDF), where we observe that the Top-10 experts (10 out of 64) receive almost $75\%$ tokens, leading to routing imbalance in MoE training.
    If the experts are evenly distributed among GPUs as in expert parallelism~\cite{GShard}, such imbalanced workloads would result in severe resource under-utilization, as most experts need to wait for the slowest.
    \item \textbf{Smoothness and continuousness}. In Figure~\ref{fig:dynamic}, we present evolution of expert loads throughout the entire training process, where different intervals represent different experts. We observe that the load of each expert is continuously changing during training, for example, from less to more, from more to less, and from more to less and then more again, etc, which poses routing fluctuation when training MoE models.
    Fortunately, the load of each expert does not change dramatically in a short period of time, which means a smooth and continuous change.
\end{itemize}

\textbf{Challenges.}
Motivated by these observations, our work explores how to develop a system that achieves workload balance by adjusting the expert-to-device mapping (i.e., $\mathcal{P}$) for better system efficiency while maintaining the model quality. Moreover, the system should dynamically adapt to the varying routing distribution during the training process. However, since both token routing and expert loads are decided by the data-dependent gating mechanism, we can not determine the optimal mapping ahead of its execution. The main challenge lies in designing and modifying the expert-to-device mapping efficiently under the fast GPU computation and rapid change in workloads. Another challenge is how to efficiently implement these irregular MoE operations.


\textbf{Opportunities.} Fortunately, as previously mentioned, the distribution changes smoothly and continuously. This means that the optimal expert-to-device mapping would not shift significantly in a short period of time. Therefore, it is feasible to refine the mapping based on the ad-hoc routing determination, without the need to predict the optimum for the next few training steps. Furthermore, the cost of computation and communication can be estimated before the actual execution. Based on the cost models, we can predict the benefits and overheads of different mapping candidates to find the best one.

\begin{figure}[t]
  \includegraphics[width=0.7\linewidth]{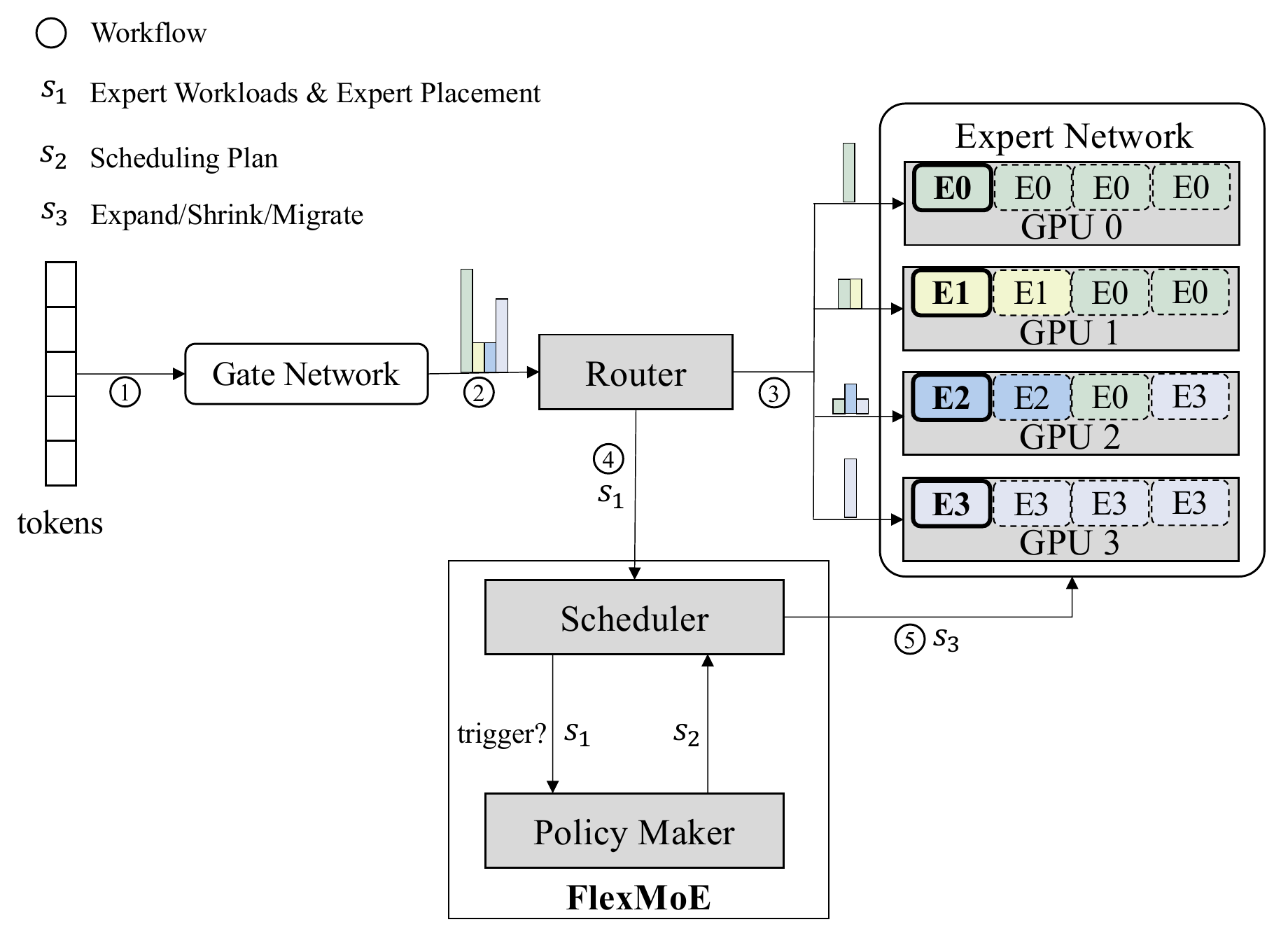}
  \caption{\oursys{} System Architecture.}
  \label{fig:system}
\end{figure}

\section{\oursys{} Design}
\label{sec:design}

\subsection{{Overview}}
The workflow of our \oursys{} is illustrated in Figure~\ref{fig:system}, where (1) input tokens are first processed by the gate network to determine their corresponding target experts and then (2) the router leverages a greedy algorithm to distribute tokens to different replicas of each expert.
Finally, (3) each replica performs computations on its assigned tokens and delivers the output results back.

To handle the dynamic workload imbalance, 
we have designed a dynamic expert management 
mechanism that adaptively adjusts the expert-to-device mapping $\mathcal{P}$ and the assignment of tokens $\mathcal{I}$ during training.
In addition to the regular workflow, we also introduce two new components: \texttt{Scheduler} and \texttt{Policy Maker}.
(4) \texttt{Scheduler} monitors the real-time loads of experts and sends them to \texttt{Policy Maker} if the current imbalance metric (i.e., balance ratio in Equation~\ref{equ:balance_ratio}) exceeds the predetermined threshold. 
Based on the received loads and the current placement of experts, \texttt{Policy Maker} produces modification instructions and sends them back to \texttt{Scheduler}.
\texttt{Scheduler} then interacts with the executor of current DL frameworks and triggers modifications of expert placement at runtime.

To cope with the large decision space of flexible expert-to-device mapping and its dynamic modifications,
\oursys{} proposes the abstraction of \vexpert{} to represent the minimum unit for scheduling, as detailed in Section~\ref{sec:vexpert}.
Moreover, \oursys{} decouples the expert placement modifications using three placement modification primitives, including \texttt{Expand}, \texttt{Shrink} and \texttt{Migrate}, as explained in Section~\ref{sec:scheduler}. 
Finally, we design a greedy heuristic algorithm to generate a sequential combination of these modification primitives to produce the adjustment plan of expert-to-device mapping, as described in Section~\ref{sec:policy}.

\vspace{-3mm}
\subsection{Dynamic Expert Management and \vexpert{}}
\label{sec:vexpert}

To tackle the optimization problem related to expert-to-device mapping and its corresponding token routing, we introduce a novel abstraction called \vexpert{}, which defines the minimum unit for scheduling GPU computations for an expert and enables dynamic expert management. 
The \vexpert{} abstraction helps in determining how to duplicate experts, when to increase or decrease replicas, and how to partition tokens between replicas.

\vexpert{} has the following characteristics: 
\begin{itemize}[leftmargin=*]
    \item Each vExpert can only be assigned as the replica of exactly one expert and process part of tokens for the master expert.
    \item Each vExpert shares weights with other vExperts for the same expert on the same GPU, which means packing the same vExperts of the same GPU. 
    \item Evenly workloads (i.e., tokens) partitioning among all vExperts for the same expert. 
\end{itemize}
With this abstraction, we can significantly reduce the large search space of the optimization problem by making decisions at the \vexpert{} level.
Besides, we have designed three placement modification primitives to manage the dynamic expert management at the \vexpert{} level, including \texttt{Expand}, \texttt{Shrink} and \texttt{Migrate} (see Section~\ref{sec:scheduler} for details). These primitives allow for arbitrary modifications of expert placement by composing them in different ways.
Moreover, because the amount of data in each iteration is fixed ($\sum B_i = B$), we can calculate the ideal capacity of each vExpert ($B/(G*E)$) as a reference for decision-making, where $B$ is the total number of input tokens, $G$ is the number of GPUs and $E$ is the number of \vexpert{s} on each device. 

\setlength{\textfloatsep}{0.1cm}
\begin{algorithm}[t]
    \small
	\SetAlgoLined
	\SetKwProg{Fn}{Function}{}{end}
    	\SetKwInput{KwData}{Input}
	\SetKwInput{KwResult}{Output}
	\KwData{$\mathcal{I}$: the assignment of tokens;\\ \quad\quad\quad$\mathcal{P}$: the expert-to-device mapping\;} 
	\SetInd{0.61em}{0.61em}
    \While{is training}{
    $balance\_ratio$ $\gets$ \texttt{balance}($\mathcal{I}$, $\mathcal{P}$) \ // Equation~\ref{equ:balance_ratio} \;
    \While{$balance\_ratio$ $>$ threshold}{
        $plan$ $\gets$ \textit{MakeSchedulingPlan}($\mathcal{I}$, $\mathcal{P}$) \ // Algorithm~\ref{alg:policy_maker} \;
        \If{$plan$ is empty}{
            break \;
        }
        $\mathcal{P}$ $\gets$ Modify Expert Placement $\mathcal{P}$ with $plan$ \;
        $balance\_ratio$ $\gets$ \texttt{balance}($\mathcal{I}$, $\mathcal{P}$) \;
    }
	$\mathcal{P}$ $\gets$ Modify Expert Placement $\mathcal{P}$ with \texttt{Migrate} \;
	}
\caption{Scheduler: Monitor Real-Time Workloads}
\label{alg:scheduler}
\end{algorithm}

\subsection{Scheduler}
\label{sec:scheduler}
Figure~\ref{fig:system} illustrates the role of the \texttt{Scheduler} in connecting the \texttt{Runtime Executor} and the \texttt{Policy Maker} in \oursys{}. 
Our design includes a metric, the balance ratio, which measures real-time workloads and determines when to trigger the \texttt{Policy Maker} for adjustment decisions. 
Additionally, we propose three atomic primitives for the \texttt{Runtime Executor} to use in scheduling experts. These primitives will be further explained in the following section.

\textbf{Balance ratio.}
Owing to the synchronous execution mode in MoE layer, the slowest GPU will dominate the finish time of all-to-all communication as well as the global training. Thus, we design the balance ratio as Equation~\ref{equ:balance_ratio}, which adds up the loads of all \vexpert{s} on a single GPU and finds the max value among all GPUs to get the final results.
While there may be other metrics that can be used to measure loads, we compare the balance ratio in Equation~\ref{equ:balance_ratio} with variance as a metric and provide a detailed analysis of our findings in Section~\ref{sec:ablation_study}.

\begin{align}
\texttt{balance}(\mathcal{I}, \mathcal{P}) = \frac{1}{\sum_{(e,g) \in \mathcal{P}} I_{e,g}/|\mathcal{G}|} \mathop{\max}_{g \in \mathcal{G}}\sum_{(e,g) \in \mathcal{P}} I_{e,g}
\label{equ:balance_ratio}
\end{align}

\textbf{Expand.} When an expert receives increasing workloads, previously allocated \vexpert{} resources may handle them slower than others and thus produce stragglers (over-utilized). In this scenario, \oursys{} will call the \texttt{Expand} primitive to allocate an extra new \vexpert{} resource for this expert at a time. 
The \texttt{Expand} primitive copies the expert parameter as well as the corresponding optimizer states from the source \vexpert{} to the target \vexpert{} via parameter sharing for intra-GPU communication and point-to-point communication of inter-GPU communication. 

\textbf{Shrink.} In the opposite, when previously allocated \vexpert{} resources are not fully utilized due to decreasing workloads (under-utilized), \oursys{} will call the \texttt{Shrink} primitive to release an \vexpert{} resource at a time. The \texttt{Shrink} primitive is executed by a tag without any communication, and thus introduces no overheads.

\textbf{Migrate.} 
If an expert retains multiple replicas on different GPUs, the training is slowed down due to the need for extra communications (i.e., gradients all-reduce) for synchronizations.
The efficiency of this communication depends on the number of GPUs involved and their locality. 
To reduce the number of GPUs holding expert replicas and lower the synchronization cost, \oursys{} will call the \texttt{Migrate} primitive exchanges the model states between two \vexpert{s}.

We present the workflow of \texttt{Scheduler} in Algorithm~\ref{alg:scheduler},
which monitors the real-time workloads, i.e., the assignment of tokens $\mathcal{I}$, and measures the balance ratio under current placements $\mathcal{P}$ (Line 1-2). If current balance ratio exceeds the pre-defined $threshold$, \texttt{Scheduler} will repeatedly ask the \texttt{Policy Maker} for modification primitives until no beneficial modifications exist (Line 3-8). Then, \texttt{Scheduler} turns to the \texttt{Migrate} operation to reduce the synchronization cost and continuously optimizes it at backend (Line 9).
Intuitively, the \vexpert{}-based \texttt{Expand} and \texttt{Shrink} primitives tackle the complicated expert-device mapping problem under the dynamic changing workloads, and the \texttt{Migrate} primitive continuously optimizes the placement of replicas for each expert.

\setlength{\textfloatsep}{0.1cm}
\begin{algorithm}[t]
    \small
	\SetAlgoLined
	\SetKwProg{Fn}{Function}{}{end}
	\SetKwInput{KwData}{Input}
	\SetKwInput{KwResult}{Output}
	\KwData{$\mathcal{I}$: the assignment of tokens;\\ \quad\quad\quad$\mathcal{P}$: the expert-to-device mapping\;
	}
	\SetInd{0.61em}{0.61em}
	\Fn{MakeSchedulingPlan($\mathcal{I}$, $\mathcal{P}$):}{
    	Estimate time cost $t_0$ with $\mathcal{I, P}$ by Equation~\ref{equ:problem_formulation}\;
    	\For{$e \in \mathcal{E}$}{
    	  $n_e \gets $ number of \vexpert{} allocated for $e$ \;
    	  $cap_e \gets \mathcal{I}_e / n_e $ // capacity of \vexpert{} for $e$ \;
    	}
        $e_0,\ e_1 \gets \argmax_{e \in \mathcal{E}} \{cap_e\},\ \argmin_{e \in \mathcal{E}} \{cap_e\}$ \;
        Estimate time cost $t_1$ after expanding $e_0$ and shrinking $e_1$\;
        \If{$t_1 < t_0$}{
            \Return \{(\texttt{Expand}, $e_0$), (\texttt{Shrink}, $e_1$)\} \;
        }
        \Return \{\ \} \;
    }
   
\caption{Policy Maker: \vexpert{}-Based Scheduling}
\label{alg:policy_maker}
\end{algorithm}

\subsection{Policy Maker}
\label{sec:policy}
Using the modification primitives described above, \oursys{} employs an efficient \vexpert{}-based scheduling algorithm for generating sequential modification operations, as shown in Algorithm~\ref{alg:policy_maker}.
Specifically, we leverage a cost-model driven search planning approach, which makes decisions based on feed-backs from simulating the training time.

To model the training cost of an MoE layer, we decompose it into three parts as Equation~\ref{equ:problem_formulation}, including computation cost $T_C$, All-To-All communication cost $T_{A2A}$ and expert synchronization cost $T_{Sync}$.
For each part, we build cost models that take into account input variables (e.g., $\mathcal{I}$: the assignment of tokens and $\mathcal{P}$: the expert-to-device mapping) as well as environmental variables (e.g., TPS: tokens-per-second for an expert). 
By leveraging a profiling-based approach, we first profile the function's running time under different input sizes and then estimate the corresponding environmental variables. 
Besides, we also consider the cost of expert adjustments, which could be executed concurrently with model training.

\textbf{Computation Cost.} The computation cost refers to the time taken for an expert to perform the forward and backward computation of experts during training, which can be formulated as: 

\begin{align}
T_C(I_{e,g}) &= \frac{I_{e,g}}{\text{$TPS$}} 
\label{equ:comp_cost}
\end{align}

\noindent where $I_{e,g}$ represents the number of input tokens received by expert $e$ on GPU $g$, and $TPS$ represents the throughput (tokens per second) of given GPUs to calculate an expert, which is obtained from profiling. 
The computation cost is proportional to the number of input tokens received by an expert, and inversely proportional to the throughput of the GPU. In other words, the more input tokens an expert receives, the longer it will take to compute the forward and backward passes, and the slower the GPU, the longer it will take to process each token.

\textbf{All-To-All Cost.} The All-To-All operation is involved to send tokens to target experts and send back their results after processing, which will be called for totally 4 times in each training step.
We predict the All-To-All cost by a topology-aware model as Equation~\ref{equ:a2a_cost}, where the expert $e$ on GPU $g$ has received $I_{e,g}.count(g^\prime)$ tokens from GPU $g^\prime$ and $Bw_{g, g^\prime}$ is the profiled bandwidth between GPU $g$ and GPU $g^\prime$. Our cost model considers the intra-node bandwidth (e.g., PCIe, NvLink) and inter-node bandwidth (e.g., IB, NIC) separately.
\begin{align}
    T_{A2A}(I_{e,g}) = 4 * \sum_{g^\prime \in \mathcal{G}}{
    \frac{I_{e,g}.count(g^\prime)}{Bw_{g, g^\prime}}
    }
    \label{equ:a2a_cost}
\end{align}

\textbf{Synchronization Cost.} 
When expert $e$ holds multiple replicas on different GPUs, their gradients must be synchronized via AllReduce communication, which is determined by the message size, the number of involved devices and their connected network. 
We enumerate different device groups and profile them before training to get their BPS (bytes-per-second). We predict the synchronization cost for the expert $e$ by finding the device group that includes $e$ (i.e., $\mathcal{P}.index(e)$) and then obtaining its corresponding $BPS$ from profiling data. ${size}(e)$ represents the size of gradients for an expert.

\begin{align}
    T_{Sync}(\mathcal{P}, e) &= \frac{{size}(e.gradients)}{BPS(\mathcal{P}.index(e))}
    \label{equ:sync_cost}
\end{align}

\textbf{Expert Adjustment Cost.} 
\oursys{} proposes three modification primitives, including \texttt{Expand}, \texttt{Shrink} and \texttt{Migrate}. The \texttt{Expand} and \texttt{Migrate} primitives involve transferring model states from the source GPU to the target GPU via NCCL Point-to-Point communication. On the other hand, the \texttt{Shrink} primitive is executed with no overheads by marking a tag. 
We predict the cost of transferring model states as $\frac{{size(e.model\_states)}}{Bw_{g,g^\prime}}$.

The scheduling policy of \texttt{Policy Maker} is summarized in Algorithm~\ref{alg:policy_maker}. Firstly, the algorithm gets the time cost $t_0$ of current placement as the baseline (Line 2), and then finds the expert id with the maximum workload and the expert id with the minimum workload (Line 3-6).
After that, our \texttt{Policy Maker} estimates the time cost $t_1$ after applying the \texttt{Expand} and \texttt{Shrink} primitives (Line 7) and decides whether to return the modification by comparing $t_0$ and $t_1$ (Line 8-10).

\setlength{\textfloatsep}{0.1cm}
\begin{algorithm}[t]
    \small
	\SetAlgoLined
	\SetKwProg{Fn}{Function}{}{end}
	\SetKwInput{KwData}{Input}
	\SetKwInput{KwResult}{Ouptut}
	\KwData{$\mathcal{I}$: Expert Workloads; $\mathcal{P}$: Expert Placement; $g$: Current GPU\;}
	\KwResult{$r$: Routing plans}
	\SetInd{0.61em}{0.61em}
	/*iterate over all experts*/ \;
	\For{e $\in$ $\mathcal{E}$}{
	$n_e \gets $ number of \vexpert{} allocated for $e$ \;
	$cap_e \gets \mathcal{I}_e / n_e$ // capacity of \vexpert{} for $e$ \; 
	$r_{e,g} \gets \min(cap_e \times n_{e,g},\ \mathcal{I}_{e,g})$ // locality first \;
	$s_e \gets \mathcal{I}_{e,g} - r_{e,g}$ // tokens for other GPUs \;
	$c_{e,g} \gets r_{e,g} - \mathcal{I}_{e,g}$ // local available capacity \;
    \For{$g^\prime \in$ $\mathcal{G} - \{g\}$}{
        $c_{e, g^\prime} \gets \min(cap_e \times n_{e, g^\prime},\ \mathcal{I}_{e, g^\prime}) - \mathcal{I}_{e, g^\prime}$ \;
        $r_{e, g^\prime} \gets s_e \times c_{e, g^\prime}/\sum c_e$ // proportional to availability \;
    }
	}
	\Return $r \gets \{r_{e}$ | $e \in \mathcal{E}\}$ \;
\caption{Flexible Token Routing}
\label{alg:routing_plan}
\end{algorithm}

\section{Implementation}
\label{sec:impl}
We have implemented the proposed mechanisms and algorithms on the top of PyTorch~\cite{pytorch19} by adding new customized operators and CUDA kernels. 
\oursys{} is also a part of a novel distributed DL system Hetu~\cite{scis2022hetu,miao2021het,miao2022hetgmp}.
To schedule dynamic dataflow more efficiently, \oursys{} incorporates the following system-level optimizations:

\textbf{Flexible Token Routing.} 
\texttt{Router} should efficiently transfer input tokens from multiple devices (e.g., GPUs) to their target experts according to the complicated expert-to-device mapping $\mathcal{P}$.
To accomplish this, \oursys{} has designed a greedy policy as Algorithm~\ref{alg:routing_plan}, which prefers to route tokens to the local GPU and then scatters the remaining tokens to other GPUs in
proportion to their available capacity.
\oursys{} also implements a efficient expert-wise layout transformation to arrange the inputs in a continuous space for each expert.

\textbf{Paralleled Operation Modification.}
\oursys{} employs a queue to sequentially insert modification primitives triggered by the \texttt{scheduler}, such as \texttt{Expand}, \texttt{Shrink} and \texttt{Migrate}.
To reduce the time cost of adjustments and kernel launch, we merge several consecutive and parallelizable operations to run them concurrently.
For example, if two operations share the same source and destination, they can  be merged to increase the message size and improve the bandwidth utilization. Meanwhile, if they share neither source or destination, they can be executed concurrently to improve the utilization of clusters.

\textbf{AllReduce Coordination.} 
When the \vexpert{s} of a single GPU are assigned to different experts, it should call the synchronizations separately for each expert, and may cause deadlock since the order of calls is inconsistent for different GPUs~\cite{jeaugey2017nccl}. 
To avoid this deadlock problem, we assign a logical id to each expert and the logical id of each replica (\vexpert{}) is same as its main expert. Then, each GPU invokes synchronizations in ascending order of experts' logical id. 

\textbf{Best-Effort Adjustment.} 
Since the placement modifications may block the training process, it is not clear whether executing the current modification will be beneficial due to the dynamic workloads.
To address this problem, \oursys{} leverages a separate CUDA stream to conduct adjustments concurrently with the available network bandwidth and adopts the best-effort adjustment to avoid hindering the training process. the \texttt{Scheduler} interacts with the DL executor to determine whether to call the first operation in the candidate queue.

\textbf{NCCL Group Management.} 
\oursys{} adopts NCCL~\cite{jeaugey2017nccl} library to perform collective communication among GPUs and multiple NCCL groups are required to execute the complicated synchronization of experts.
However, there is a maximum number of live NCCL groups that can remain, and it is inefficient to eliminate the groups once they have been utilized.
\oursys{} employs a \textit{Least Recently Used (LRU)} cache to maintain nccl groups and therefore reduces the costs of group creations and eliminations.

\begin{table}[t]
\small
    \caption{Models for Evaluation.}
    \vspace{-2mm}
    \begin{center}
        \begin{tabular}{c|cccccc}
        \hline 
        \textbf{Model}  & \textbf{Params.}  & \textbf{\#Layer}  & $\mathbf{d_{Model}}$ & $\mathbf{d_{FFN}}$ & \textbf{\#Expert}\\ 
        \hline
        \hline
        BERT-MoE-S  & 0.988B & 12 & 768 & 3072 & 32\\
        BERT-MoE-L  & 6.69B & 24 & 1024 & 4096 & 64\\
        GPT-MoE-S  & 0.988B & 12  & 768 & 3072 & 32\\
        GPT-MoE-L  & 39B & 24 & 2048 & 8192 & 64\\
        \hline
        Swin-MoE-S  & 946M & 24  & - & - & 32\\
        Swin-MoE-L  & 1.83B & 24  & - & - & 64\\
        \hline
        \end{tabular}
    \end{center}
\label{tab:model_structure}
\end{table}

\vspace{-3mm}
\section{Evaluation}
\label{sec:eval}
In this section, we present the detailed evaluation results to demonstrate the effectiveness and scalability of \oursys{}.
\vspace{-3mm}
\subsection{Experimental Setup}
\textbf{Machine environment.} We conduct experiments on Azure 
VMs~\cite{azure_a100}, each equipped with 192-core 
AMD CPUs and 8 NVIDIA Ampere A100 GPUs.
The GPUs are connected via NVLink 3.0 intra-node
and the servers are connected via 8 InfiniBand NICs (8*200 Gbps totally). RDMA is used by default and the PyTorch version is 1.11.

\textbf{Baselines.} We compare \oursys{} with other popular frameworks, including DeepSpeed~\cite{rajbhandari2022deepspeed} and FasterMoE~\cite{he2022fastermoe}. Deepspeed used expert parallelism, which was first proposed by GShard~\cite{GShard}.
FasterMoE proposed the shadowing strategy to replicate the popular expert among all GPUs. 

\textbf{Benchmarks and datasets.}
Our evaluations are conducted by scaling representative transformer models in different application domains with the MoE architecture, including BERT~\cite{bert} and GPT~\cite{gpt2} in NLP and Swin~\cite{liu2021swin} in CV, as shown in Table~\ref{tab:model_structure}. 
We adopt Top-2 Gate for each model, which is adopted by widely used sparse MoE models (GShard~\cite{GShard}, GLaM~\cite{du2021glam}, V-MoE~\cite{vmoe}), and set the capacity factor as $1.0$ for each expert and balance loss as $0.001$. 
We pretrain BERT-MoE with masked language modeling (MLM) and next-sentence-prediction (NSP)~\cite{bert} tasks, GPT-MoE with language modeling task (LM)~\cite{gpt2}, Swin-MoE with image classification task~\cite{liu2021swin,tutel}. 
NLP models are trained on wikipedia~\cite{wikipedia} and vision models are trained on the ImageNet-1K~\cite{imagenet}. All the hyper-parameters (e.g., learning rate) are fixed as for the same model.

\begin{table*}[t]
    \small
    \caption{Comparison on model quality}
    \vspace{-2mm}
    \begin{center}
        \begin{tabular}{c|c|cc|cc|c|cc}
        \hline 
        \multirow{3}{*}{}  & \multirow{3}{*}{Metric} & \multicolumn{2}{c|}{Masked LM} & \multicolumn{2}{c}{Language Modeling} \vline & \multirow{3}{*}{Metric} & \multicolumn{2}{c}{Image Classification} \\
        \cline{3-6} \cline{8-9}
        &~ & BERT-  & BERT-  &  GPT- & GPT- & ~ & Swin- &  Swin-\\
        &~ & MoE-S & MoE-L  &  MoE-S & MoE-L & ~ & MoE-S &  MoE-L \\
        \hline
        \hline
        \multirow{2}{*}{DeepSpeed} & \multirow{2}{*}{PPL$\ \downarrow$} & \multirow{2}{*}{3.53} & \multirow{2}{*}{3.31} & \multirow{2}{*}{12.2} & \multirow{2}{*}{10.71} & acc$@$1$\ \uparrow$ & 77.316 & 77.022 \\
        & & & & & & acc$@$5$\ \uparrow$ & 93.838 & 93.642
        \\
        \hline
        \multirow{2}{*}{\oursys{}} & \multirow{2}{*}{PPL$\ \downarrow$} & \multirow{2}{*}{3.14} & \multirow{2}{*}{3.07} & \multirow{2}{*}{11.72} & \multirow{2}{*}{10.47} & acc$@$1$\ \uparrow$ &  77.754 & 77.109 \\
        & & & & & & acc$@$5$\ \uparrow$ & 94.042 & 93.663
        \\
        \hline
        \end{tabular}
    \end{center}
\label{tab:model_perf}
\vspace{-2mm}
\end{table*}

\begin{figure}[t]
    \centering
        \subfigure[32 GPUs]{
            \scalebox{0.95}{\includegraphics[width=0.5\columnwidth]{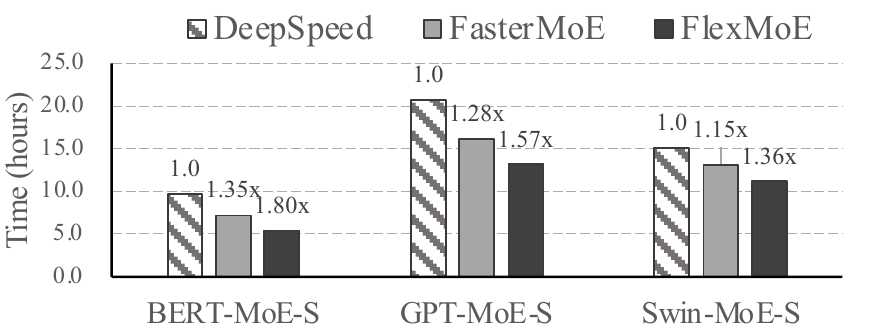}}
            \label{fig:e2e_32gpus}
        }
        \subfigure[64 GPUs]{
            \scalebox{0.95}{\includegraphics[width=0.5\columnwidth]{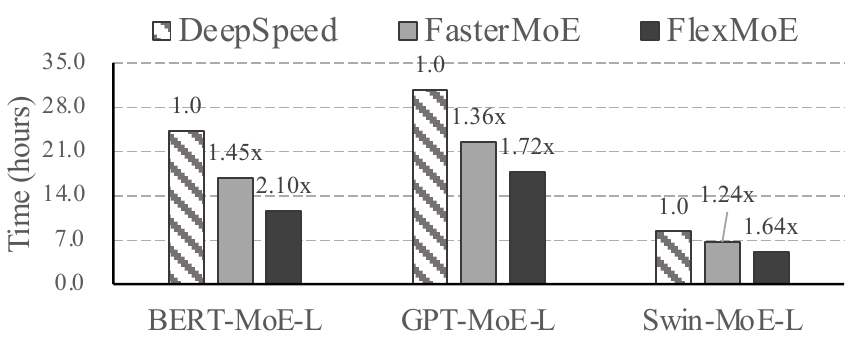}}
            \label{fig:e2e_64gpus}
        }
    \vspace{-3mm}
    \caption{Comparison on system efficiency}
    \label{fig:e2e}
\end{figure}

\subsection{Overall Performance}
To evaluate the end-to-end efficiency of \oursys{}, we compare it with DeepSpeed and FasterMoE. 
For DeepSpeed, we use the traditional expert parallelism approach and set the number of experts per GPU as 1 in every MoE layer. 
We evaluate two different width of models, X-MoE-S with 32 experts and X-MoE-L with 64 experts, on 32 GPUs and 64 GPUs, respectively.

\textbf{Model Quality.} 
To demonstrate the importance of not dropping tokens, we compare the model quality of \oursys{} with DeepSpeed on various models and various tasks. 
We use the validation perplexity for language models (e.g., BERT-MoE and GPT-MoE), which is the lower the better, and top-1/top-5 accuracy of Imagenet-1K for vision models (e.g., Swin-MoE). 
As shown in Table~\ref{tab:model_perf}, \oursys{} outperforms DeepSpeed in almost all tasks, indicating that the limited capacity in existing MoE systems (e.g., DeepSpeed) can lead to model quality degradation. 
Considering the training cost, the Swin-MoE models (scaled based on the Swin-B model) are trained for 100 epochs (less than 300 epochs in the standard configuration) and thus the accuracy is slightly worse than the benchmark, where $94.04\%$ of \oursys{} v.s. $96.5\%$ of the benchmark as for top-5 accuracy. And we believe it's enough to show the benefits of no dropping tokens. What's more, Swin-MoE-L performs slightly worse than Swin-MoE-S as the size of training data is small and thus it may lead to overfitting.

\textbf{System Efficiency.} We also evaluate \oursys{} against other SOTA systems on efficiency. To measure efficiency, we record the required training time to achieve the target model quality, and the results are illustrated in Figure~\ref{fig:e2e}. 
Our experiments show that \oursys{} achieves the best training efficiency, outperforming DeepSpeed by $1.70\times$ on average and up to $2.10\times$, and FasterMoE by $1.30\times$ on average and up to $1.45\times$.
As mentioned above, although DeepSpeed obtains the smallest iteration time thanks to its limited capacity, it drops tokens to skip the expert network and thus requires more iterations to converge. 
FasterMoE proposes the dynamic shadowing strategy for loading balance, which replicates the popular expert on all GPUs. However, due to its coarse-grained expert management (i.e., on 1 GPU or on all GPUs), it falls back to a sub-optimal solution. 
As the number of GPUs increases, FasterMoE suffers from the global synchronization of expert replicas.

In addition to the expert networks, the time of models' training also consists of other parts, such as non-experts' computation, optimizers' update and communication. As \oursys{} only optimizes the execution of the expert networks, we will mainly focus on analyzing the MoE layer in the following sections. 
\vspace{-3mm}

\subsection{Ablation Study}
\label{sec:ablation_study}
To verify the effectiveness of \oursys{}, we conduct several ablation studies, as demonstrated below.

\begin{figure*}[t]
    \centering
        \subfigure[Different metrics ]{
            \scalebox{0.72}{\includegraphics[width=0.5\columnwidth]{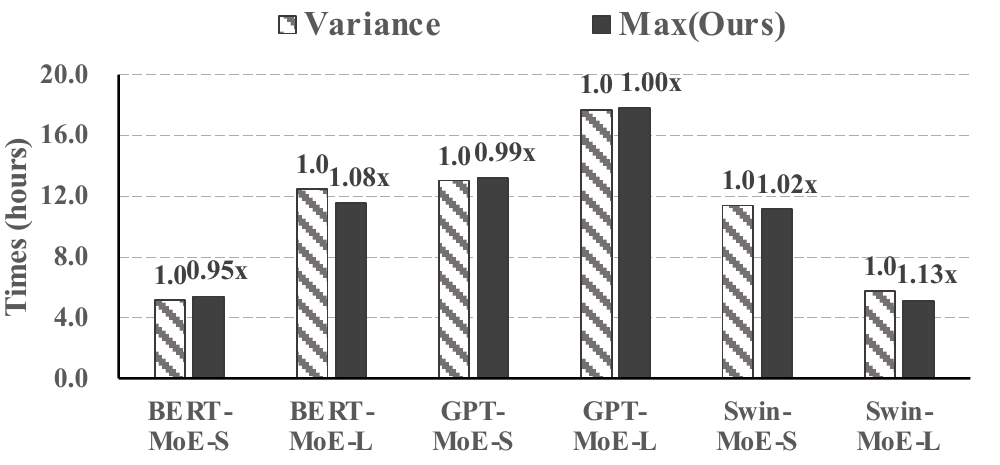}}
            \label{fig:different_metrics}
        }
        \subfigure[Different Scheduling Polices]{
            \scalebox{0.6}{\includegraphics[width=0.5\columnwidth]{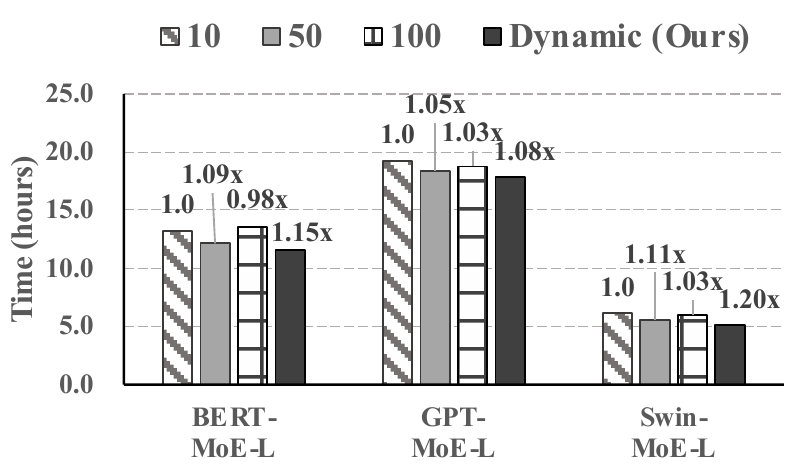}}
            \label{fig:different_polices}
        }
        \subfigure[Cost Models]{
            \scalebox{0.55}{\includegraphics[width=0.5\columnwidth]{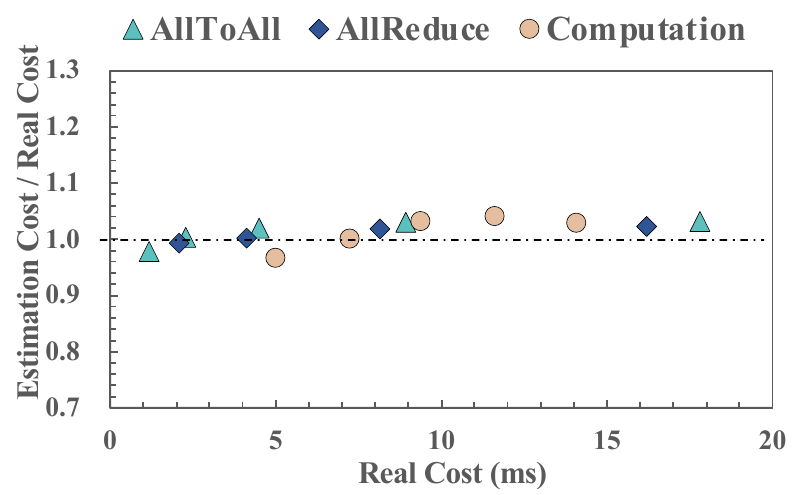}}
            \label{fig:cost_models}
        }
    \vspace{-3mm}        
    \caption{Study of different metrics, different scheduling policies and cost models}
\end{figure*}


\textbf{Different Metrics.} \oursys{} utilizes the balance ratio as a key metric to trigger adjustments. To investigate the impact of alternative metrics, , we also study the use of \textit{Variance} as a ratio, formulated as $\sum_{g \in \mathcal{G}}(I_g - \overline{I})^2 / {|\mathcal{G}|}$, in addition to the \textit{Max(ours)} ratio given by Equation~\ref{equ:balance_ratio}.
Results are presented in Figure~\ref{fig:different_metrics} and show that \textit{Max(ours)} outperforms \textit{Variance} by $1.03\times$ on average and up to $1.13\times$ on Swin-MoE-L. 
Meanwhile, \textit{Variance} also performs well on BERT-MoE-S and GPT-MoE-S. 
Because the training time is dominated by the slowest expert, \textit{Max(ours)} is a simple but effective metric. \textit{Variance} takes the global workload distribution into consideration, which triggers adjustment more frequently but often gets empty operations from \texttt{Policy Maker} as it is not always relevant to the actual training.

\textbf{Different Scheduling Policies.} \oursys{} dynamically triggers adjustment according to the pre-defined balance ratio. We also conduct experiments on static scheduling policies, which triggers the adjustments in the fixed interval and executes them completely before training. As shown in Figure~\ref{fig:different_polices}, our dynamic scheduling outperforms both large interval and small interval because of the dynamic changing workloads. 
{Specifically, small interval will trigger adjustments frequently and thus introduce more adjustment costs, while large interval can not tackle the dynamic workloads well because it can not make adjustments in time.}
Our dynamic policy dynamically decides the adjustments based on current workloads, which is suitable for the dynamic workloads.

\textbf{Evaluation on Cost Models}
Figure~\ref{fig:cost_models} demonstrates the effectiveness of our cost models, where we compare the estimation cost to real cost on different input sizes for computation/alltoall/allreduce respectively. It can be observed that our estimation results are very close to the real execution costs for all experimental models, where the average prediction error is less than $3\%$.

\subsection{Token Efficiency and Expert Efficiency}
\begin{figure}[t]
    \centering
        \subfigure[Token efficiency and Expert efficiency]{
            \scalebox{0.37}{\includegraphics[width=\columnwidth]{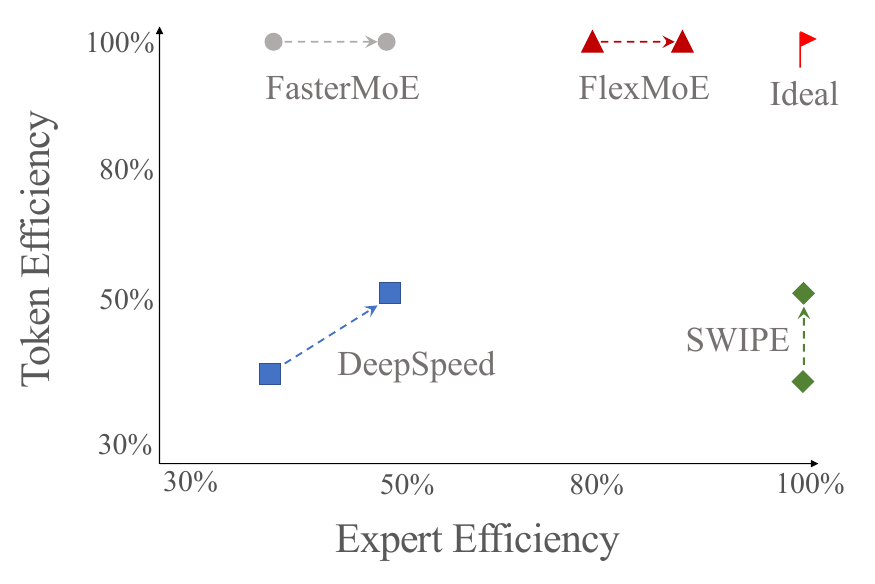}}
            \label{fig:token_and_expert_efficiency}
        }
        \subfigure[Scalability]{
            \scalebox{0.4}{\includegraphics[width=\columnwidth]{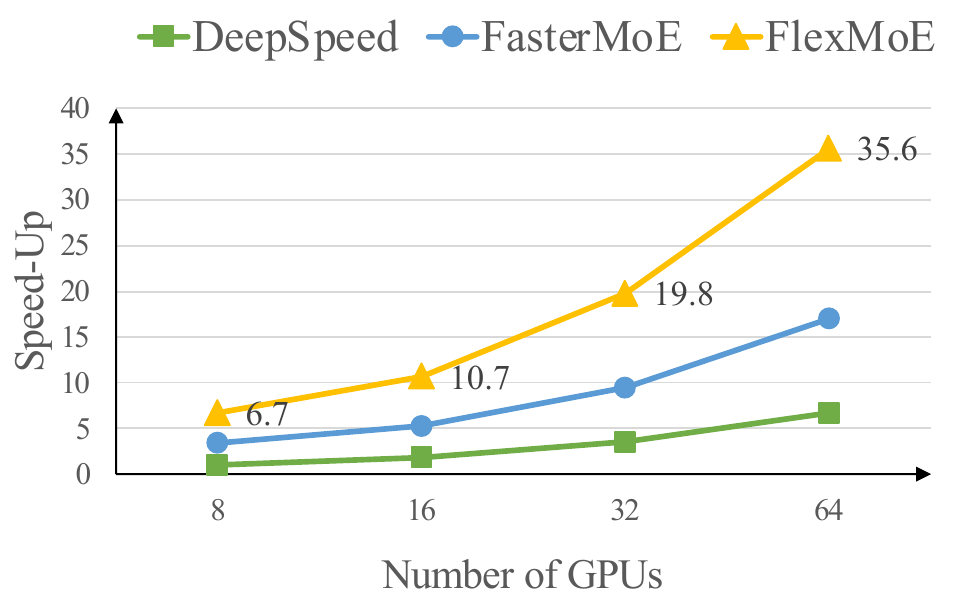}}
            \label{fig:scalability}
        }
    \vspace{-3mm}
    \caption{Figure~\ref{fig:token_and_expert_efficiency} shows the trends of token efficiency and expert efficiency for different methods during training. And Figure~\ref{fig:scalability} shows the scalability for different methods.}
\end{figure}

In this section, we analyze both \textit{token efficiency} and \textit{expert efficiency} for different MoE training methods during the whole training process. 
\textit{Token efficiency} refers to the fraction of input tokens that are processed by the expert network and 
\textit{expert efficiency} refers to the meaningful computation of GPUs. $100\%$ of both metrics is the ideal status of an MoE layer, shown as the red flag in Figure~\ref{fig:token_and_expert_efficiency}.
The traditional expert-parallel method (e.g., DeepSpeed) obtains low token efficiency and expert efficiency as it drops tokens beyond capacity for loading balance. SWIPE, proposed by BaGuaLu~\cite{ma2022bagualu}, improves expert efficiency by modifying the gating algorithm to re-assign inputs to other experts for strict load balance. 
However, this approach changes the relations between tokens and experts, thus leads to low token efficiency.
FasterMoE~\cite{he2022fastermoe} replicates hot experts on each GPU and guarantees no tokens dropping in the implementation. However, it doesn't take load balance as its design goals and obtains low expert efficiency. Our system, \oursys{}, guarantees $100\%$ token efficiency and optimizes the allocation of computation resources to balance the workloads among GPUs, and is the closest to the ideal. With the training progressing, the imbalanced workloads are getting better due to the punishment of balance loss and all methods are moving towards to better efficiency.

\subsection{Scalability}
We also evaluate the scalability of \oursys{} on 8, 16, 32 and 64 GPUs, which are conducted on a single MoE layer with 64 experts. The results are presented in Figure~\ref{fig:scalability}, which are normalized to the throughput of DeepSpeed-8GPUs.  
Results show \oursys{} significantly outperforms DeepSpeed and FasterMoE. 
As experiments are conducted on a high-speed interconnected cluster (8*300Gbps intra-node and 8*200 Gbps inter-node), balanced computation among GPUs plays a key role and \oursys{} is targeting as it.

\section{Conclusion}
\label{sec:conclusion}
In this paper, we presented \oursys{}, a novel solution to address the dynamic imbalanced challenges encountered during the training of large-scale MoE models.
By integrating a scheduling module on top of existing DNN frameworks, \oursys{} monitors data traffic, creates scheduling plans, and dynamically adjusts the expert-to-device mapping during training. Our empirical results on six popular MoE models demonstrate that \oursys{} outperforms DeepSpeed by an average of $1.70\times$ and up to $2.10\times$, while also surpassing FasterMoE by an average of $1.30\times$ and up to $1.45\times$.

\section{Acknowledgments}
This work is supported by the National Key Research and Development Program of China (No. 2020AAA0105200), the National Natural Science Foundation of China (No. 61832001 and U22B2037) and PKU-Tencent joint research Lab. 
Bin Cui is the corresponding author.

\nocite{tinyscript, peng2021graph, davoudian2021workload, ge2021watuning, wang2023global, DBLP:journals/pvldb/FuJSC19, DBLP:journals/pvldb/FuMJXC22}
\bibliographystyle{ACM-Reference-Format}
\bibliography{sample-base}

\end{document}